\documentclass[12pt,thmsa]{article}
\usepackage{amsfonts}
\usepackage{amssymb}
\usepackage{graphicx}

\input{tcilatex}
\begin{document}

\begin{center}
{\large Information Entropy of Two Cooper Pair Boxes Interacting with an
Environment }

{\Large \ }~

Heba Kadry Abdel-Hafez and Nordin Zakaria

{\footnotesize HPC Service Center, Universiti Teknologi PETRONAS, \\
               31750 Tronoh, Perak, Malaysia \\
               Email: hkadry1@yahoo.com, nordinzakaria@petronas.com.my}
\end{center}

\bigskip

Adopting the framework of two-coupled superconducting charges model,
we discussed the information entropy of two qubits initially
prepared in a mixed state and allowed to interact with their
environment. The impact of the different parameters of the system is
explicitly investigated. Present calculations show that a strong
mutual entropy is obtained for a long interaction time when the two
qubits are initially in a mixed state with the environment switched
off. We have identified a class of two-qubit states that have a rich
dynamics when the deviation between the characteristic energies of
the qubits become minimum. It was found that the total correlation
decreases abruptly to zero in a finite time due to the influence of
the strong environment.


\section{Introduction}

In designing and analyzing cryptosystems and protocols, mathematical
concepts are critical in supporting the claim that the intended cryptosystem
is secure \cite{chu08} and different concepts have important impact in this
direction, e.g. the pseudo entropy as a measure of information content of
ionic state due to ion-laser interaction in a single trapped ion has been
used \cite{oba04} superconductor-based quantum information processor \cite%
{ast06}, and optical properties of quantum charge qubits structures are
currently in the focus of the research activity owing to their promising
potentiality in different areas of present-day science and technology \cite%
{nis07,you05}. One of the physical realizations of a solid-state qubit is
provided by a Cooper pair box which is a small superconducting island
connected to a large superconducting electrode, a reservoir, through a
Josephson junction \cite{gri07}. Superconducting charge qubits (Cooper pair
boxes) are a promising technology for the realization of quantum computation
on a large scale \cite{coo04,nak99,pas05,pas03}.

On the other hand, one of the major challenges in the field of quantum
information theory is to get a deep understanding of how local operations
assisted by classical communication performed on a multi-level quantum
system can affect the entanglement between the spatially separated systems.
Despite a lot of progress in the last few years, it is still not fully
understood. For instance, even for the simple question of whether a given
state is entangled or not, no general answer is known \cite{doh04}. An
interesting question raised in \cite{guh04} is whether there is any
relationship between the uncertainty principle and entanglement or not.
Recently, a general definition of entropy squeezing for a two-level system
has been presented \cite{fan00} and showed that the information entropy is a
measure of the quantum uncertainty of atomic operators. Also, the
number-phase entropic uncertainty relation for the multiphoton coherent
state and nonlinear coherent state is studied and compared with an ordinary
coherent state \cite{jos01}.\bigskip

The purpose of this paper is to propose the quantum mutual entropy
as a measure of the total correlation between two-coupled
superconducting charge qubits, each qubit is based on a Cooper pair
box connected to a reservoir electrode through a Josephson junction.
Our approach in relating the correlation of the pair of two charge
qubits system to the maximum entangled state generated has the merit
of directly involving quantities having a clear physical meaning.
Applying our criterion to the considered system we are able to fully
exploit the novelty of our point of view to explain the ability to
generate maximum entangled states and the corresponding correlation.

This paper is organized as follows: in Section 2, we will describe the
Hamiltonian of the system of interest, and obtain the explicit analytical
solution of the master equation describing the dynamics of two qubits in the
presence of phase decoherence. In Section 3, we discuss the total
correlation of the system by virtue of the mutual entropy in the absence or
presence of the decoherence. Finally, Section 4 presents the conclusions and
an outlook.

\section{The model}

We consider two charge qubits and couple them by means of a miniature
on-chip capacitor \cite{pas05,ave85}. The read-out of each qubit, in this
case, is done similar to the single qubit read-out and connect a probe
electrode to each qubit. External controls that we have in the circuit are
the dc probe voltages $V_{b_{1}}$ and $V_{b_{2}}$, $d_{c}$ gate voltages $%
V_{g_{1}}$ and $V_{g_{2}}$, and pulse gate voltage $V_{p}$. The information
on the final states of the qubits after manipulation comes from the
pulse-induced currents measured in the probes. By doing routine current--
voltage--gate voltage measurements, we can estimate the capacitances. We
then perform state manipulation and demonstrate qubit--qubit interaction.
The Hamiltonian of the system in the charge representation can be written as
\cite{pas05}%
\begin{eqnarray}
\hat{H} &=&\hbar \sum_{n_{1}=0}^{\infty }\sum_{n_{2}=0}^{\infty }\eta
_{1}(n_{1},n_{2})|n_{1},n_{2}\rangle \langle n_{1},n_{2}|  \label{ham1} \\
&&-\frac{E_{J_{1}}}{2}\left( |n_{1},n_{2}\rangle \langle
n_{1}+1,n_{2}|+|n_{1},n_{2}+1\rangle \langle n_{1}+1,n_{2}+1|\right)
\nonumber \\
&&-\frac{E_{J_{2}}}{2}\left( |n_{1},n_{2}\rangle \langle
n_{1},n_{2}+1|+|n_{1}+1,n_{2}\rangle \langle n_{1}+1,n_{2}+1|\right) ,
\nonumber
\end{eqnarray}
The parameter $\eta _{1}(n_{1},n_{2})=E_{c_{1}}\left( n_{g_{1}}-n_{1}\right)
^{2}+E_{c_{2}}\left( n_{g_{2}}-n_{2}\right) ^{2}+E_{m}\left(
n_{g_{1}}-n_{1}\right) \left( n_{g_{2}}-n_{2}\right) .$ Here, $n_{1}$ and $%
n_{2}$ ($n_{1},n_{2}=0,\pm 1,\pm 2,...$) define the number of excess
Cooper pairs in the first and the second Cooper pair boxes respectively, and $%
n_{g_{1,2}}=(C_{g_{1,2}}V_{g_{1,2}}+C_{p}V_{p})/2e$ are the normalized
charges induced on the corresponding qubit by the $d_{c}$ and pulse gate
electrodes. The eigenenergies, $E_{k}$ $(k=0,1,2,...$), of the Hamiltonian in (%
\ref{ham1}) form $2e$-periodic energy bands corresponding to the ground ($%
k=0 $), first excited ($k=1$), etc. states of the system. The energy bands $%
E_{k} $ for the one-dimensional case were first introduced in Ref. \cite%
{pas05,ave85}. $E_{c_{1}}$, $E_{c_{2}}$ and $E_{m}$ give the characteristic
energies of Cooper pair of the first qubit, Cooper pair charging energy of
the second qubit and the coupling energy, respectively. $E_{c1,2}=\frac{%
4e^{2}C_{\varepsilon 2,1}}{2(C_{\varepsilon 1}C_{\varepsilon 2}-C_{m}^{2})},$
and $E_{m}=\frac{4e^{2}C_{m}}{C_{\varepsilon 1}C_{\varepsilon 2}-C_{m}^{2}},$%
where $C_{\varepsilon _{2,1}}$ are the sum of all capacitances connected to
the corresponding Cooper pair box including the coupling capacitance $C_{m}$
and $e$ is the electron charge.

If the circuit is fabricated to have the following relation between the
characteristic energies: $E_{J_{1,2}}\sim $ $E_{m}<E_{c_{1,2}}$, then one
can use a four-level approximation for the description of the system ($%
|e_{1}e_{2}\rangle ,|g_{1}e_{2}\rangle ,|e_{1}g_{2}\rangle $ and $%
|g_{1}g_{2}\rangle $) around $n_{g_{1}}=n_{g_{2}}=0.5$ while other charge
states are separated by large energy gaps. These four charge states can be
used as a new basis for the Hamiltonian (\ref{ham1}).

Decoherence is not always due to the interaction with an
environment, but it may also be due, to the fluctuations of some
classical parameter or internal variable of a system. This is a
different form of decoherence, which is present even in isolated
systems which is called non-dissipative
decoherence. In this paper, we follow the standard procedure \cite%
{bre02,lid03,gar00} to write the time evolution of the system density
operator $\hat{\rho}(t)$ in the following form
\begin{equation}
\frac{d}{dt}\hat{\rho}(t)=-\frac{i}{\hbar }[\hat{H},\hat{\rho}]-\frac{\gamma
}{2\hbar ^{2}}[\hat{H},[\hat{H},\hat{\rho}]],  \label{mas}
\end{equation}%
where $\gamma $ is the phase decoherence rate. Equation (\ref{mas}) reduces
to the ordinary von Neumann equation for the density operator in the limit $%
\gamma \rightarrow 0.$ The equation with the similar form has been proposed
to describe the intrinsic decoherence \cite{mil91}. Under Markov
approximations the \ solution of the master equation can be expressed as
follows%
\begin{equation}
\rho (t)=\left(
\begin{array}{c}
\rho _{ee,ee} \\
\rho _{eg,ee} \\
\rho _{ge,ee} \\
\rho _{gg,ee}%
\end{array}%
\begin{array}{c}
\rho _{ee,eg} \\
\rho _{eg,eg} \\
\rho _{ge,eg} \\
\rho _{gg,eg}%
\end{array}%
\begin{array}{c}
\rho _{ee,ge} \\
\rho _{eg,ge} \\
\rho _{ge,ge} \\
\rho _{gg,ge}%
\end{array}%
\begin{array}{c}
\rho _{ee,gg} \\
\rho _{eg,gg} \\
\rho _{ge,gg} \\
\rho _{gg,gg}%
\end{array}%
\right) .
\end{equation}
We use the notation $|ij\rangle =|i_{1}\rangle \otimes |j_{2}\rangle ,$ $%
(i,j=e,g),$ where $|e_{1(2)}\rangle $ and $|g_{1(2)}\rangle $ are the basis
states of the first (second) qubits and $\rho _{ij,lk}(t)=\langle ij|\rho
(t)|lk\rangle $ corresponds the diagonal ($ij=lk$) and off-diagonal ($ij\neq
lk)$ elements of the final state density matrix $\rho (t).$ From here on,
for tractability of notation and without loss of generality, we denote by $%
\rho _{ij}(t)=\rho _{ij,ij}(t),$ the probability of finding the two-coupled
qubits in the charge state $|ij\rangle .$\bigskip

\section{Information entropy}

In the theory of open system or the reduction theory, one often considers
two subsystem $\mathcal{H}_{1}$ and $\mathcal{H}_{2}$ represented by Hilbert
space. Let $\mathfrak{S}(\mathcal{H}_{i}),\,\,(i=1,2)$ be state spaces (the
set of all density operators). Also $\mathfrak{S}(\mathcal{H}_{1}\otimes
\mathcal{H}_{2})$ denotes the state space in the composite system $\mathcal{H%
}_{1}\otimes \mathcal{H}_{2}$ \cite{yu06}. The following decomposed
states in composite system are called disentangled states :
\begin{equation}
|\Psi \rangle =|\psi _{1}\rangle \otimes |\psi _{2}\rangle ,{\ \ \ \ \ \ }%
|\psi _{1}\rangle \in \mathfrak{S}(\mathcal{H}_{1}),{\ \ \ \ \ \ }|\psi
_{2}\rangle \in \mathfrak{S}(\mathcal{H}_{2}).
\end{equation}%
However in general it is almost impossible to decompose the states in
composite system like the above, that is, the following states exist :
\begin{eqnarray}
|\Psi \rangle &=&\alpha |\psi _{1}\rangle \otimes |\psi _{2}\rangle +\beta
|\phi _{1}\rangle \otimes |\phi _{2}\rangle ,  \nonumber \\
&&|\psi _{1}\rangle ,{\ \ \ }|\phi _{1}\rangle \in \mathfrak{S}(\mathcal{H}%
_{1}),\,\qquad \,|\psi _{2}\rangle ,{\ \ \ \ }|\phi _{2}\rangle \in
\mathfrak{S}(\mathcal{H}_{2})  \nonumber \\
&&|\alpha |^{2}+|\beta |^{2}=1,{\ \ \ \ \ \ \ \ }\alpha \neq 0,{\ \ \ \ \ \ }%
\beta \neq 0.
\end{eqnarray}%
The above states which can not be described by product states of two
subsystem, are called entangled states. For the entangled states $\mathcal{E}%
_{t}^{\ast }\rho \in \mathfrak{S}(\mathcal{H}_{1}\otimes \mathcal{H}_{2})$,
the quantum mutual entropy is defined by the following formula as a distance
(difference) from a disentangled state $tr_{\mathcal{H}_{1}}\varrho \otimes
tr_{\mathcal{H}_{2}}\varrho \in \mathfrak{S}(\mathcal{H}_{1}\otimes \mathcal{%
H}_{2})$
\begin{equation}
I_{\mathcal{E}_{t}^{\ast }\rho }\left( \rho ^{A},\rho ^{B}\right) =tr%
\mathcal{E}_{t}^{\ast }\rho (\log \mathcal{E}_{t}^{\ast }\rho -\log (tr_{%
\mathcal{H}_{1}}\mathcal{E}_{t}^{\ast }\rho \otimes tr_{\mathcal{H}_{2}}%
\mathcal{E}_{t}^{\ast }\rho )).
\end{equation}%
Note that if the entangled state $\mathcal{E}_{t}^{\ast }\rho $ is a pure
state, $S(\mathcal{E}_{t}^{\ast }\rho )=0$ and then $S(tr_{\mathcal{H}_{2}}%
\mathcal{E}_{t}^{\ast }\rho )=S(tr_{\mathcal{H}_{1}}\mathcal{E}_{t}^{\ast
}\rho )$ by Araki-Lieb inequality \cite{ara70}, thus we have $I_{\mathcal{E}%
_{t}^{\ast }\rho }\left( \rho ^{A},\rho ^{B}\right) =2S(tr_{\mathcal{H}_{2}}%
\mathcal{E}_{t}^{\ast }\rho )$.

\smallskip We now suppose that the initial state of the Cooper pairs is a
mixed state:
\begin{equation}
\rho =\cos ^{2}\left( \frac{\xi }{2}\right) \left\vert
e_{1}e_{2}\right\rangle \langle e_{1}e_{2}|+\sin ^{2}\left( \frac{\xi }{2}%
\right) \left\vert g_{1}g_{2}\right\rangle \langle g_{1}g_{2}|.  \label{2}
\end{equation}%
In what follows, we are going to derive a general form of the quantum mutual
entropy $I_{\mathcal{E}_{t}^{\ast }\rho }\left( \rho ^{A},\rho ^{B}\right) $
for two charge qubits. Using equations (\ref{fdens}) and (\ref{2}), one can
write the von-Neumann entropy of the two-qubit state as
\begin{equation}
S(\mathcal{E}_{t}^{\ast }\rho )=-\sum_{i=1}^{4}\lambda _{i}^{(AB)}(t)\log
_{2}\lambda _{i}^{(AB)}(t),  \label{entropy01}
\end{equation}%
where $\lambda _{1}^{(AB)}(t)=\langle e_{1}e_{2}|\mathcal{E}_{t}^{\ast }\rho
|e_{1}e_{2}\rangle ,$ $\lambda _{2}^{(AB)}(t)=\langle g_{1}g_{2}|\mathcal{E}%
_{t}^{\ast }\rho |g_{1}g_{2}\rangle ,$%
\begin{eqnarray*}
\lambda _{3,4}^{(AB)}(t) &=&\frac{\langle e_{1}g_{2}|\mathcal{E}_{t}^{\ast
}\rho |e_{1}g_{2}\rangle +\langle g_{1}e_{2}|\mathcal{E}_{t}^{\ast }\rho
|g_{1}e_{2}\rangle }{2} \\
&&\pm \frac{1}{2}\sqrt{\left( \langle e_{1}g_{2}|\mathcal{E}_{t}^{\ast }\rho
|e_{1}g_{2}\rangle +\langle g_{1}e_{2}|\mathcal{E}_{t}^{\ast }\rho
|g_{1}e_{2}\rangle \right) ^{2}+4|\langle e_{1}g_{2}|\mathcal{E}_{t}^{\ast
}\rho |g_{1}e_{2}\rangle |^{2}}.
\end{eqnarray*}%
The von Neumann entropy of reduced density operator of the first qubit $%
S(\rho _{t}^{A})$ is given
\begin{equation}
S(\rho _{t}^{F})=-\sum_{i=1}^{2}\lambda _{i}^{(A)}(t)\log _{2}\lambda
_{i}^{(A)}(t).  \label{entropy02}
\end{equation}%
where $\lambda _{1}^{(A)}(t)=\langle e_{1}|\rho ^{(A)}|e_{1}\rangle ,$ $%
\lambda _{2}^{(A)}(t)=\langle g_{1}|\rho ^{(A)}|g_{1}\rangle ,$ while the
von Neumann entropy of reduced density operator of the second qubit $S(\rho
_{t}^{B})$ can be written as
\begin{equation}
S(\rho _{t}^{F})=-\sum_{i=1}^{2}\lambda _{i}^{(B)}(t)\log _{2}\lambda
_{i}^{(B)}(t).
\end{equation}%
where $\lambda _{1}^{(B)}(t)=\langle e_{2}|\rho ^{(B)}|e_{2}\rangle ,$ and $%
\lambda _{2}^{(B)}(t)=\langle g_{2}|\rho ^{(B)}|g_{2}\rangle .$

Thus we rigorously obtain the analytical expression of the quantum mutual
entropy of the system under consideration in the following form (\ref%
{entropy01}) and (\ref{entropy02})
\begin{eqnarray}
I_{\mathcal{E}_{t}^{\ast }\rho }\left( \rho ^{A},\rho ^{B}\right)  &\equiv
&tr\mathcal{E}_{t}^{\ast }\rho (\log \mathcal{E}_{t}^{\ast }\rho -\log (\rho
_{t}^{A}\otimes \rho _{t}^{B}))  \nonumber \\
&=&S(\rho _{t}^{(A)})+S(\rho _{t}^{(B)})-S(\mathcal{E}_{t}^{\ast }\rho )
\nonumber \\
&-&\sum_{i=1}^{2}\lambda _{i}^{(A)}(t)\log _{2}\lambda
_{i}^{(A)}(t)-\sum_{i=1}^{2}\lambda _{i}^{(B)}(t)\log _{2}\lambda
_{i}^{(B)}(t)  \nonumber \\
&&+\sum_{i=1}^{4}\lambda _{i}^{(AB)}(t)\log _{2}\lambda _{i}^{(AB)}(t).
\label{dem}
\end{eqnarray}%
Throughout this paper, we use the quantum mutual entropy equation (\ref{dem}%
) as a measure of the total correlation of two qubits. \textrm{It
turns out to be rather easy to derive an analytic expression for the
quantum mutual entropy for any given system, since }with the help of
equation (\ref{dem}) it is possible to study the total correlation
of any two-qubit system when the system starts from a mixed state.
This measure is not only a function of interaction time but also a
function of coupling energy, characteristic energies and initial
state setting. We now turn to the task of numerically analyzing the
effect of the system parameters on the total correlation and
occupation probabilities of the four charge states. It is obvious
that correlation does not appear when the coupling energy parameter
$E_{m}=0$ because the two qubits state becomes separable and
correlation is not generated. For $E_{m}\neq 0$, we found that
correlation is produced between the qubits regardless of
characteristic energies when the qubits are initially prepared in
mixed state and $\gamma =0$ (see Figure 1).
\begin{figure}[tbph]
\begin{center}
\includegraphics[width=12cm,height=8cm]{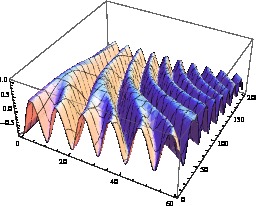}
\end{center}
\caption{Numerical results for quantum mutual entropy as a function of the
scaled time and coupling energy. The parameters used in these figures are $%
E_{J1}=E_{J2}=30,\protect\gamma =0,$ and the initial state of the two qubits
is maximally entangled. }
\end{figure}

It is shown that the qubits exist between the upper and lower states only
with very small probability of being in the intermediate states and maximum
entangled mixed states in the form $\rho =\frac{1}{2}\left( \left\vert
e_{1}e_{2}\right\rangle \langle e_{1}e_{2}|+\left\vert
g_{1}g_{2}\right\rangle \langle g_{1}g_{2}|\right) $ is obtained at some
instances that corresponding to the maximum correlation, i.e. given enough
time, the system will therefore reaches a state where both excited and
ground states have equal occupation probabilities. Indeed in the limit that $%
\theta \sim 0,$ the quantum mutual entropy is only twice of the quantum von
Numann entropy. In the general case ($i.e.,$ $\theta \neq 0$), the final
state does not necessarily become a pure state, so that we need to make use
of $I_{\mathcal{E}_{t}^{\ast }\rho }\left( \rho _{t}^{A},\rho
_{t}^{F}\right) $ in the present model. Thus our initial setting enables us
to discuss the variation of the quantum mutual entropy for different values
of the mixed state parameter $\theta .$ \textrm{A related model allowing an
analytic treatment of the quantum entanglement as well as valuable insight,
namely an ensemble of two identical qubits }coupled to a cavity field\textrm{%
\ }has been discussed in \textrm{\ \cite{aty04}. }
\begin{figure}[tbph]
\begin{center}
\includegraphics[width=12cm,height=6cm]{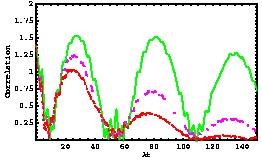}
\end{center}
\caption{The time evolution of the quantum mutual entropy as a function of
the scaled time and different values of the decoherence parameters, where, $%
\protect\gamma =0.01$ (solid curve), $\protect\gamma =0.1$ (dashed curve), $%
\protect\gamma =0.5$ (dot curve). The other parameters are the same as
figure 1.}
\end{figure}

Since one of our main goals is to analyze the different possible
types of behavior in the coupled qubits dynamics, we have to
identify the relevant parameters that determine the different
behavior regimes. In Figure 2, we consider the quantum mutual
entropy as a function of the scaled time for different values of the
decoherence parameter $\gamma $. This case is quite interesting
because the quantum mutual entropy function oscillates around the
maximum and minimum values for small period of the interaction time
and the local maximum decreases as time goes on. We have shown here
a new phenomena where the periodic oscillations occur irrespective
of the decoherence. A slight change in $\gamma $ therefore,
dramatically alters the quantum mutual entropy. It should be noted
that for a special choice of the decoherence parameter, the
situation becomes interesting where, correlation decreases abruptly
to zero in a finite time due to the influence of the strong
environment. Nevertheless, the sharp peaks lie within the region
between the two maximum values of the correlation occurring in a
similar way for different values of the decoherence parameter but
the local maximum is decreased, such that with higher $\gamma $
these oscillations of the quantum mutual entropy decay to local
minimum which is close to zero. From our further calculations, we
notice that if the difference between the two characteristic
energies is substantially larger than the coupling energy, the
effect of the decoherence on the total correlation dynamics
diminishes rapidly. Then, we have emphasized and identified that for
some higher values of $\gamma $ there were no persisting periods
found to lie between the two maximum values. Similar effects have
been observed in different systems experimentally \cite{alm07} and
theoretically \cite{ebe07}.

From our further calculations, we see that the time evolution of
total correlation for different values of the characteristic
energies $E_{J_{1}}$ and $E_{J_{2}}$ when the charge qubits are
initially prepared in a mixed state keeping the coupling energy
fixed, the number of oscillations of the quantum mutual entropy is
increased as the deviation between $E_{J_{1}}$ and $E_{J_{2}}$ is
increased while its dynamics exhibits the same qualitative behaviors
which has been observed in Figure 1. However, it is interesting to
note that when characteristic energies take different values i.e. $%
E_{J_{1}}\neq E_{J_{2}}$ even with different values of the mixed
state parameter $\theta $, the intermediate level always populated
and oscillate around small values. Total correlation is maximized
when the qubits are initially prepared in a mixed state and the
deviation takes its minimum value ($E_{J_{1}}\simeq E_{J_{2}})$.
Compared with Figure 1, the long levied maximum correlation is no
longer exists in this case and increasing the deviation between the
characteristic energies produces more oscillations from the early
stage of interaction with sharp peaks of the local maximum of the
correlation function. Because of different values of the
characteristic energies which are control the qubits, the
interaction becomes more complicated and brings about faster
oscillations in quantum mutual entropy.

Our quantitative results are of potential use in the analysis of a broad
class of relevant experimental situations dealing with quantum information
theory. It is worth mentioning that the dynamics of coupled superconducting
charges systems has always been of interest, but has recently attracted even
more attention because of application in quantum computing. Several systems
have been suggested as physical realizations of quantum bits allowing for
the needed control manipulations, and for some of them the first elementary
steps have been demonstrated in experiments \cite{nak99}.

\section{Conclusion}

Summarizing, we have analyzed the quantum correlation of a physically
interesting system interacting with its environment in the context of
two-coupled superconducting charges model. We have explicitly evaluated the
exact solution of the density matrix of the system and worked out the
effects of different parameters of the system as well as environmental
parameter on the dynamics of the system. In particular, a considerable
enhancement of maximum entangled mixed state generation of semiconductor
qubits is obtained when even weak coupling energy is employed. We have
identified the relation between the quantum mutual entropy as a measure of
quantum correlation between the two charge qubits and maximum entangled
state generation. This seems significant, and one then wonders whether the
trend might continue with the general multi-level case. With an increase in
exposure to the environment, and for small values of the coupling energy, we
found faster decays of the total correlation when the qubits are initially
prepared in mixed state. This fast decay rate of quantum correlation is a
generic feature in a variety of physical processes where decoherence is
important. This kind of numerical investigation constitutes the first
quantitative characterization of the relation between the total correlation
and maximum entangled state generation for the superconducting charges
qubits.


\begin{thebibliography}{99}
\bibitem{chu08} Y. F. Chung, Z. Y. Wub and T. S. Chen, Information Sciences
178 (2008) 2044--2058

\bibitem{oba04} A.-S.F. Obada, S. Furuichi, H.F. Abdel-Hameed and M.
Abdel-Aty, Information Sciences 162 (2004) 53--61

\bibitem{ast06} O. Astafiev, Yu. A. Pashkin, Y. Nakamura, T. Yamamoto, and
J. S. Tsai, Phys. Rev. Lett. 96, 137001 (2006)

\bibitem{nis07} A. O. Niskanen, K. Harrabi, F. Yoshihara, Y. Nakamura, S.
Lloyd, and J. S. Tsai Science, \textbf{316}, 723 (2007)

\bibitem{you05} J. Q. You and F. Nori, Physics Today, 58, 42 (2005); J. Q.
You, J. S. Tsai, and F. Nori, Phys. Rev. B \textbf{73}, 014510 (2006)

\bibitem{gri07} E. J. Griffith, C. D. Hill, J. F. Ralph, H. M. Wiseman and
K. Jacobs, Phys. Rev. B \textbf{75}, 014511 (2007)

\bibitem{coo04} K. B. Cooper, M. Steffen, R. McDermott, R. W. Simmonds,
S.Oh, D. A. Hite, D. P. Pappas, and J. M. Martinis, Phys. Rev. Lett. \textbf{%
93}, 180401 (2004).

\bibitem{nak99} Y. Nakamura, Y. A. Pashkin, and J. S. Tsai, Nature. \textbf{%
398}, 786 (1999).

\bibitem{pas05} Yu. A. Pashkin, T. Yamamoto, O. Astafiev, Y. Nakamura, D. V.
Averin, T. Tilma, F. Nori and J. S. Tsai, Physica C \textbf{426--431} 1552
(2005).

\bibitem{pas03} Yu. A. Pashkin, T. Yamamoto, O. Astafiev, Y. Nakamura, D. V.
Averin, J. S. Tsai, Nature \textbf{421}, 823 (2003).

\bibitem{doh04} A. C. Doherty,  P. A. Parrilo and F. M. Spedalieri, Phys.
Rev. A \textbf{69}, 022308 (2004)

\bibitem{guh04} O. Guhne  and M. Lewenstein, Phys. Rev. A \textbf{70},
022316 (2004)

\bibitem{fan00}  M. F. Fang, P. Zhou and S. Swain, J. Mod. Opt. \textbf{47},
1043 (2000)

\bibitem{jos01} A. Joshi, J. Opt. B: Quantum Semiclass. Opt. \textbf{3}, 124
(2001)

\bibitem{ave85} \textrm{D. V. Averin, A.B. Zorin and K. K. Likharev, JETP
\textbf{61, }407 (1985).}

\bibitem{bre02} H.-P. Breuer and F. Petruccione, The theory of open quantum
systems, Oxford University Press, Oxford, (2002)

\bibitem{lid03} D. A. Lidar and K. B. Whaley, in Irreversible quantum
dynamics edited F.Benatti and R. Floreanini, Spring Lecutre Notes in
Physics,Vol. 62, Berlin (2003), p.83.

\bibitem{gar00} C. W. Gardiner and P. Zoller, Quantum Noise
(Springer-Verlag, Berlin, 2000).

\bibitem{mil91} G.J. Milburn, Phys. Rev. A \textbf{44}, 5401 (1991); S.
Schneider and G.J. Milburn, Phys. Rev. A \textbf{57}, 3748 (1998); S.
Schneider and G.J. Milburn, Phys. Rev. A \textbf{59}, 3766 (1999).

\bibitem{yu06} T. Yu and J. H. Eberly, Phys. Rev. Lett. \textbf{97}, 140403
(2006); \textit{ibid} Quantum Information and Computation, (2007) in press

\bibitem{ara70} H. Araki and E. Lieb, Commun. Math. Phys. \textbf{18}, 160
(1970)

\bibitem{aty04} M. Abdel-Aty and A.-S. F. Obada, J. Math. Phys. \textbf{45},
4271 (2004)

\bibitem{alm07} M. P. Almeida, F. de Melo, M. Hor-Meyll, A. Salles, S. P.
Walborn, P. H. Souto Ribeiro, L. Davidovich, Science, \textbf{316}, 579
(2007)

\bibitem{ebe07} J. H. Eberly and T. Yu, Science, \textbf{316}, 555 (2007);
A.-S. F. Obada and M. Abdel-Aty, Phys. Rev. B, \textbf{75}, 195310 (2007)
\end{thebibliography}
\end{document}